\documentclass[aps,prl,showpacs,twocolumn,floats]{revtex4}
\usepackage{amssymb}

\usepackage{graphicx}
\usepackage{bm}

\begin{document}

\bibliographystyle{prsty}
\author{E. M. Chudnovsky}
\affiliation{Physics Department, Lehman College, The City
University of New York \\ 250 Bedford Park Boulevard West, Bronx,
New York 10468-1589, U.S.A.}
\date{\today}

\begin{abstract}
We study the dependence of the intrinsic spin Hall effect on the
crystal symmetry and geometry of experiment. The spin current is
obtained and the Hall voltage caused by the polarization of the
electron spins is computed. The unique dependence of the effect on
the crystal symmetry permits the choice of geometry in which the
spin Hall effect can be unambiguously distinguished from the
effects due to the orbital motion of charge carriers and due to
the magnetic field generated by the transport current.
\end{abstract}

\title{Intrinsic Spin Hall Effect in Non-Cubic Crystals}

\maketitle

The spin Hall effect consists of opposite ``Hall'' currents for
charge carriers with opposite spin polarizations. Remarkably, no
magnetic field is needed for the electrons with spin up and spin
down to veer in opposite directions. The effect was initially
observed in GaAs strips by Kato et al. via Kerr signal
\cite{Kato}, and by Wunderlich et al. via polarization of light
emitted at the recombination of spin-polarized electron-hole pairs
\cite{Wunderlich}. When numbers of spin-up and spin-down electrons
are equal, no net electric current across the strip is generated
by the spin Hall effect. If, however, the electrons are polarized
due to, e.g., injection from a ferromagnet \cite{DasSarma}, then
the numbers of electrons veering left and right are different and
a Hall voltage is produced. Valenzuela and Tinkham used this
effect to measure the spin Hall conductivity in the aluminum strip
\cite{Tinkham}.

Dyakonov and Perel \cite{DP} were apparently the first to suggest
that scattering of charge carriers by unpolarized impurities in
semiconductors can lead to the spin polarization of the sample
boundaries. Their suggestion was based upon a similar effect in
atomic physics known as Mott scattering \cite{Mott}. When a
relativistic electron passes at a speed ${\bf v}$ through an atom,
the partially unscreened electric field of the nucleus, ${\bf E}$,
creates the magnetic field ${\bf B} = {\bf E} \times {\bf v}/c$ in
the coordinate frame of the electron. This field partially
polarizes the electron spin in the direction that is opposite for
the electrons passing the nucleus on the right and on the left.
Consequently, one can achieve a spatial separation of spin-up and
spin-down electrons when unpolarized electron beam passes through
unpolarized atomic target.

A number of microscopic models have been developed that extended
these ideas to solids. Various ``extrinsic'' (due to impurities)
and ``intrinsic'' (impurity-free) mechanisms have been studied
with the aim to explain quantitatively the spin Hall effect in
non-magnetic conductors \cite{Rashba}, as well as to explain the
anomalous Hall effect in magnetic materials \cite{anomalous}. Most
of the existing theoretical models are based upon the
Boltzmann-type kinetic equation that describes spin and charge
transport under certain assumptions about the collision integral.
There is also a vast amount of numerical work of disordered
systems in lattices of finite size. So far these works have not
offered any universal description of the spin Hall effect. Instead
a variety of different spin Hall effects (spin precession, side
jump, skew scattering) that are specific to the model of
spin-orbit interaction, nature of scatterers, band structure,
boundary effects, etc. has been proposed. A simple single-electron
picture of the intrinsic spin Hall effect, similar to the
treatment of the conventional Hall effect within Drude model, has
been suggested in Ref.\ \onlinecite{EC}. In this picture the Mott
scattering of the transport current by the crystal field appears
naturally within the framework of the Aharonov-Casher effect
\cite{AC}. The parameter-free expression for the spin Hall
conductivity was obtained for cubic crystals in good agreement
with experiments. In this paper we will extend the treatment of
Ref. \onlinecite{EC} to non-cubic crystals. Straightforward
experiments will be suggested that can test our predictions.

Our approach is based upon the general form of the one-electron
Hamiltonian that contains spin-orbit interaction to $1/c^2$
\cite{Drell}:
\begin{equation}\label{ham1}
{\cal{H}} = \frac{{\bf p}^2}{2m} + U({\bf r}) +
\frac{\hbar}{4m^2c^2}{\bm \sigma}\cdot\left({\bm \nabla} U \times
{\bf p}\right)\,.
\end{equation}
It is exactly this Hamiltonian, with $U({\bf r})$ being the
electric potential of the atom, that is responsible for the Mott
scattering of electrons in the atomic physics. In a solid, $U({\bf
r})$ is the electrostatic crystal potential felt by a charge
carrier. For certainty we will speak about electrons but the model
will equally apply to holes. With an accuracy to $1/c^2$
Hamiltonian (\ref{ham1}) is mathematically equivalent to \cite{AC}
\begin{equation}\label{ham2}
{\cal{H}} = \frac{1}{2m}\left({\bf p} - \frac{e}{c}{\bf
A}_{\sigma}\right)^2 + U({\bf r})
\end{equation}
where $e < 0$ is the charge of electron and \cite{EC}
\begin{equation}\label{A}
{\bf A}_{\sigma} \equiv -\frac{\hbar}{4emc}\left({\bm
\sigma}\times {\bm \nabla}U\right)\,.
\end{equation}
(We have used the mathematical fact that the action of the
operator ${\bf p}$ on ${\bf A}_{\sigma}$ is zero,
$\epsilon_{ijk}\sigma_j\nabla_i\nabla_kU = 0$.) Consequently, the
orbital motion of electrons is affected by the fictitious
spin-dependent magnetic field:
\begin{equation}\label{B}
{\bf B}_{\sigma} = {\bm \nabla} \times{\bf A}_{\sigma} = -
\frac{\hbar}{4emc}\, \left[{\bm \nabla} \times \left({\bm
\sigma}\times {\bm \nabla}U\right)\right]\,.
\end{equation}
This fictitious field produces the same effect on the orbital
motion of electrons as the real magnetic field does in the
conventional Hall effect, but with the Hall currents having
opposite directions for electrons with opposite spin
polarizations. The spin-dependent Lorentz force, ${\bf
F}_{L\sigma} = -(e/c)({\bf v}\times{\bf B}_{\sigma})$, gives rise
to the fictitious spin-dependent Hall electric field,
\begin{equation}\label{E-H}
{\bf E}_{H\sigma} = R_H \left({\bf B}_{\sigma} \times {\bf
j}_0\right)\,,
\end{equation}
where ${\bf j}_0=en{\bf v}_0$ is the transport current expressed
through the concentration $n$ and drift velocity ${\bf v}_0$ of
the electrons, and $R_H = -1/(nec)$ is the Hall constant. If the
electrons are polarized, the spin average of Eq.\ (\ref{B}) gives
rise to the effective magnetic field,
\begin{equation}\label{B-eff}
{\bf B}_{eff} =  - \frac{\hbar}{4emc}\, \langle {\bm \nabla}
\times \left({\bm \xi}\times {\bm \nabla}U\right)\rangle\,,
\end{equation}
and to the measurable Hall electric field:
\begin{equation}\label{E-eff}
{\bf E}_{H} = R_H \left({\bf B}_{eff} \times {\bf j}_0\right)\,.
\end{equation}
In these formulas $0 < \xi < 1$ is the polarization of the
electrons and $\langle ... \rangle$ denotes the space average. If
polarized electrons are injected from, e.g., a magnetic metal,
${\bf j}_0$ describes the flow of the injected electrons.

In the presence of the real magnetic field ${\bf B}$, Eq.\
(\ref{ham2}) should be replaced by
\begin{equation}\label{ham3}
{\cal{H}} = \frac{1}{2m}\left({\bf p} - \frac{e}{c}{\bf A} -
\frac{e}{c}{\bf A}_{\sigma}\right)^2 + U({\bf r}) +
\frac{g}{2}\mu_B{\bm \sigma}\cdot{\bf B}\,,
\end{equation}
where ${\bf A} = ({\bf B} \times {\bf r})/2$, $g$ is the
gyromagnetic factor, and $\mu_B$ is the Bohr magneton. Since the
definition of ${\bf A}_{\sigma}$, Eq.\ (\ref{A}), already contains
$1/c$, the cross-term proportional to ${\bf A}\cdot {\bf
A}_{\sigma}$ in Eq.\ (\ref{ham3}) has the order $1/c^3$. Within
the non-relativistic approximation of Eq.\ (\ref{ham1}), that has
accuracy to $1/c^2$, such a cross-term must be omitted.
Consequently, in the non-relativistic theory the conventional Hall
effect and the spin Hall effect are totally independent. For that
reason and in order to emphasize the consequences of the spin Hall
effect, we are considering below the case of ${\bf B} = 0$.

When writing down Eq.\ (\ref{E-H}) and Eq.\ (\ref{E-eff}) we made
an assumption that $\langle{\bf v}\times{\bf B}_{\sigma}\rangle =
\langle{\bf v}\rangle\times\langle{\bf B}_{\sigma}\rangle$. Some
justification of this assumption follows from the fact that the
trajectory of the charge carrier does not correlate strongly with
the quadrupole component of the crystal electric field contained
in the expression for ${\bf B}_{\sigma}$. Another argument is
based upon symmetry. Indeed, the only reason for $\langle{\bf
F}_{L\sigma}\rangle$ to be different from zero would be ${\bf v}_0
\equiv \langle{\bf v}\rangle \neq 0$. Consequently, $\langle{\bf
F}_{L\sigma}\rangle$ should be first order on ${\bf v}_0$. Being
perpendicular to the velocity, the force ${\bf F}_{L\sigma}$ does
not do mechanical work on the charge. Neither should $\langle{\bf
F}_{L\sigma}\rangle$ with respect to the drift motion of the
charges, rendering the form $\langle{\bf F}_{L\sigma}\rangle =
-(e/c) ({\bf v}_0\times\tilde{\bf B})$. It is natural to identify
$\tilde{\bf B}$ with ${\bf B}_{eff}$ of Eq.\ (\ref{B-eff}).

To show that space averaging in Eq.\ (\ref{B-eff}) produces a
non-zero result, one needs to compute $\langle \nabla_i\nabla_j
U({\bf r}) \rangle$. Experiments performed to date have been done
in cubic semiconductors and in aluminum that is also cubic
\cite{Kato,Wunderlich,Tinkham}. For a cubic lattice
\begin{equation}\label{cubic}
\langle{\bm \nabla}_i{\bm \nabla}_jU\rangle = C\delta_{ij}
\end{equation}
due to the cubic symmetry alone, with $C$ being a constant. This
constant can be found from the Laplace equation:
\begin{equation}\label{Laplace}
C \equiv \frac{1}{3}\langle {\bm \nabla}^2 U({\bf r}) \rangle = -
\frac{4 \pi}{3} e \langle \rho({\bf r}) \rangle\,,
\end{equation}
where $\rho({\bf r})$ is the charge density that creates $U({\bf
r})$. To make the right choice for $\rho({\bf r})$, we notice that
the spin-orbit interaction becomes larger as the electron passes
closer to the nucleus. Similar to the Mott scattering by
individual atoms, the electric neutrality of the crystal as a
whole \cite{Kravchenko} is irrelevant for our problem; the
distances that matter are the ones where the screening of the
electric charge of the nuclei is not complete \cite{EC-reply}.
Consequently, the spatial average in Eq.\ (\ref{B-eff}) must be
over short distances. Due to the periodicity of the crystal it can
be computed over the unit cell. In that sense our symmetry
argument for the cubic lattice and other lattices studied below is
similar to the argument used to compute the crystal field
(magnetocrystalline anisotropy) in magnetically ordered crystals
\cite{lectures}.

In line with the conventional approach to solids \cite{AM}, we
choose $U({\bf r})$ as the potential formed by a cubic lattice of
ions of charge $-Ze > 0$. Then $\langle \rho \rangle = -Zen_0 =
-en$ where $n_0$ and $n=Zn_0$ are concentrations of ions and
conduction electrons, respectively. This gives $C = 4\pi e^2n/3$
\cite{EC}. Those who find this argument too simplistic may want to
compare it with the approach developed by Hirsch \cite{Hirsch-99}.
In a model that replaces moving spins with stationary electric
dipoles Hirsch computed the same average over a cubic lattice of
charges numerically. To nine decimal places his result coincides
with ours up to a factor of $2$ that can be traced to the
difference in the expression for the fictitious magnetic field
\cite{Hirsch-sign}. Note that in a microscopic model the
right-hand side of Eq.\ (\ref{Laplace}) contains a sum over
delta-functions. It is therefore likely that at the microscopic
level the spin Hall effect originates from the singularity of the
Coulomb potential. Since our model does not treat the screening
effects rigorously and does not take into account the interaction
between the electrons, our result for $C$ can only be valid up to
a factor of order unity.

Substitution of Eq.\ (\ref{cubic}) into Eq.\ (\ref{B-eff}) gives
\begin{equation}\label{B-eff-C}
{\bf B}_{eff} = C\frac{\mu_B}{e^2}{\bm \xi}=\frac{4\pi}{3}n \mu_B
{\bm \xi} \,,
\end{equation}
which provides the Hall field
\begin{equation}\label{E-eff-C}
{\bf E}_{H}  = -\frac{2\pi \hbar}{3mc^2}\left({\bm \xi} \times
{\bf j}_0\right)\,.
\end{equation}
Note that this formula does not contain any dependence on the
concentration of charge carriers. Only the knowledge of ${\bm
\xi}$ and ${\bf j}_0$ are needed to compute the Hall electric
field. This allows easy analysis of experiments with polarized
electrons in cubic conductors. One can see a potential problem,
however, with interpreting the Hall voltage in Eq.\
(\ref{E-eff-C}) as a spin Hall effect. Indeed, the effective field
in Eq.\ (\ref{B-eff-C}) equals the magnetic field that the
polarized electrons would produce in a spherically shaped body;
with $4\pi/3$ being the demagnetizing factor. The magnetic field
due to polarization of electron spins in the actual sample should
be different by a factor of order unity. Nevertheless, given the
experimental uncertainties, it may be difficult to practically
distinguish the spin Hall effect described by Eq.\ (\ref{E-eff-C})
from the conventional Hall effect due to the magnetic field
produced by the electron polarization.

In fact, the above-mentioned controversy is a consequence of the
cubic symmetry. Indeed, in the conventional Hall effect only the
orbital motion of the electron matters, the electron spin is
irrelevant. On the contrary, the spin Hall effect arises from the
spin-orbit term in the Hamiltonian. Consequently, there must be a
clear way to distinguish between the two effects. As we shall see
below, the non-cubic crystals present such a possibility.
Consider, e.g., a tetragonal crystal with ${\bf n}$ being the unit
vector in the direction of the $c$-axis. By symmetry, Eq.\
(\ref{cubic}) should be now replaced with ($a = b$)
\begin{equation}\label{tetragonal}
\langle{\bm \nabla}_i{\bm \nabla}_jU\rangle = C[k_a\delta_{ij} +
(k_c-k_a) n_in_j]\,,
\end{equation}
where $k_{a} \neq k_c$ are factors of order unity. Working out the
cross-products in Eq.\ (\ref{B-eff}) one obtains
\begin{equation}\label{B-tetr}
{\bf B}_{eff} = C\frac{\mu_B}{2e^2}\left[(k_a + k_c){\bm \xi} +
(k_a - k_c)({\bf n} \cdot {\bm \xi}){\bf n}\right]\,.
\end{equation}
The new feature is the component of the effective field along the
tetragonal axis. (Notice the analogy with the magnetic anisotropy
field in a magnetically ordered uniaxial crystal \cite{lectures}.)

\begin{figure}[ht]
\begin{center}
\vspace{-4.0cm}\hspace{-0.5cm}
\includegraphics[width=90mm,angle=0]{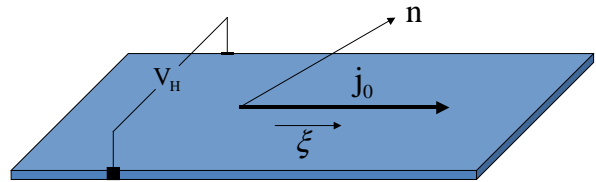}
\vspace{-4.5cm} \caption{Color online: Geometry of the proposed
experiment. The crystal anisotropy axis is at an angle with the
film. Due to the spin Hall effect the Hall voltage appears even
when the electron spins are polarized along the transport
current.} \label{geometry}
\end{center}
\vspace{0cm}
\end{figure}

Substitution of Eq.\ (\ref{B-tetr}) into Eq.\ (\ref{E-eff}) gives
\begin{eqnarray}\label{E-tetr}
{\bf E}_H & = & - \frac{\pi \hbar}{3mc^2}[(k_a +
k_c)({\bm \xi} \times {\bf j}_0) \nonumber \\
& + & (k_a-k_c)({\bf n} \cdot {\bm \xi})({\bf n}\times {\bf
j}_0)]\,.
\end{eqnarray}
This formula must be also correct for a hexagonal crystal, with
${\bf n}$ along the hexagonal axis. Its remarkable property is
that due to the second term the Hall voltage can be produced in
the sample even when the electrons are polarized along the
direction of the transport current. This would be a clear
manifestation of the Hall effect due to the spin, in contrast with
the ordinary Hall effect due to the conventional spin-independent
Lorentz force on the transport current. One possible geometry of
the experiment is shown in Fig.\ \ref{geometry}.

Until now we have studied the spin Hall effect in a system of
partially polarized electrons. Alternatively, one can study the
spin currents that would polarize the boundaries of the sample in
the absence of the polarization in the bulk. In fact, this is how
the spin Hall effect was initially observed
\cite{Kato,Wunderlich}. The spin current is defined \cite{Rashba}
as the one-particle expectation value of
\begin{equation}\label{spin-current}
j_{ik} = \frac{1}{2}e n (\sigma_i v_k + v_k \sigma_i)\,,
\end{equation}
In general it is not conserved. However, due to the relativistic
smallness of spin-orbit interaction the reversal of the electron
spin occurs only in a small fraction of scattering events.
Consequently, in a small sample at low temperature the charge
carriers typically reach the boundary of the sample before
scattering reverses their spins. In this case Eq.\
(\ref{spin-current}) provides a useful concept for the study of
the spin accumulation at the boundaries.

Since Eq.\ (\ref{spin-current}) contains Pauli matrices, the
non-zero expectation value of $j_{ik}$ is provided by the
spin-dependent part of the velocity operator. The latter is given
by \cite{EC}
\begin{equation}\label{v}
{\bf v}_{\sigma} = \frac{e\tau}{m}{\bf E}_{H\sigma}\,,
\end{equation}
where $\tau$ is the scattering time. Consider, e.g., an
orthorhombic crystal. By symmetry the principal axes of the
second-rank tensor $\langle{\bm \nabla}_i{\bm \nabla}_jU\rangle$
should be directed along the $a,b,c$ axes of the crystal. Choosing
the axes of the coordinate frame along the crystal axes, we
present $\langle{\bm \nabla}_i{\bm \nabla}_jU\rangle$ in a
diagonal form:
\begin{equation}\label{ave-ort}
\langle{\bm \nabla}_i{\bm \nabla}_jU\rangle = C k_i\delta_{ij}\,,
\end{equation}
with $k_{x,y,z}$ being generally unequal factors of order unity.
Substitution of Eq.\ (\ref{ave-ort}) into Eq.\ (\ref{B}) then
gives
\begin{equation}\label{B-ort}
B_{\sigma i} = C \frac{\mu_B}{2e^2}\sigma_i(k_x + k_y + k_z -
k_i)\,.
\end{equation}
Finally, with the help of equations (\ref{E-H}),
(\ref{spin-current}), (\ref{v}), and (\ref{B-ort}), one obtains
\begin{equation}\label{spin-current-final}
j_{ik} = \frac{\pi \hbar\sigma_0}{3mc^2}\epsilon_{ikl}(k_x + k_y +
k_z - k_i)j_{0l}\,,
\end{equation}
where $\sigma_0 = e^2n\tau/m$ is the usual charge conductivity.
According to this equation, the strength of the spin current as
compared to the charge current is determined by the factor
$\hbar\sigma_0/(mc^2)$ which also determines the ratio of spin
Hall and charge conductivities \cite{EC}. For good metals at low
temperature this ratio can be of order $10^{-4}$.

For a cubic crystal $k_x=k_y=k_z = 1$ and Eq.\
(\ref{spin-current-final}) reduces to the expression
\begin{equation}\label{j-cubic}
j_{ik} = \frac{2\pi \hbar\sigma_0}{3mc^2}\,\epsilon_{ikl}j_{0l}
\end{equation}
which is independent of the orientation of the crystal lattice
with respect to the transport current. This may present a problem
for distinguishing the polarization of the sample boundaries
generated by the spin Hall effect from the spin polarization
generated by the Zeeman effect due to the magnetic field of the
current. This controversy can be resolved by studying the spin
Hall effect in a non-cubic crystal. Consider, e.g., two
geometrically identical conducting strips, one cut along the
$aa$-plane and the other cut along the $ac$-plane of a tetragonal
crystal, with the $z$-axis being perpendicular to the plane of the
strip and the transport current being along the $y$-axis. It is
easy to see from Eq.\ (\ref{spin-current-final}) that the spin
current $j_{zx} = [{\pi \hbar\sigma_0}/(3mc^2)](k_x + k_y)j_0$,
describing the flow of $\sigma_z$ along the $x$-axis, is different
for the two strips. The ratio of these spin currents for the same
value of the transport current is given by $j_{zx}(aa)/j_{zx}(ac)=
2k_a/(k_a+k_c) \neq 1$. Consequently, the spin polarizations of
the boundaries generated by the spin Hall effect will also be
different for the two strips, while polarizations generated by the
Zeeman effect due to the magnetic field of the transport current
will be the same.

In Conclusion, we have studied the intrinsic spin Hall effect in
non-cubic crystals. The unique dependence of the effect on the
crystal symmetry permits geometry of experiment in which the spin
Hall effect can be unambiguously distinguished from the effects
caused by the orbital motion of charge carriers and by the
magnetic field of the transport current.

The author thanks David Awschalom for bringing his attention to
recent experimental papers on spin Hall effect in non-cubic
crystals \cite{Awschalom}. This work has been supported by the
Department of Energy through Grant No. DE-FG02-93ER45487.

\end{document}